# Refinement of Strategy and Technology Domains STOPE View on ISO 27001

**Heru Susanto**[123], **Fahad Bin Muhaya**[1], **Mohammad Nabil Almunawar** [2], **Yong Chee Tuan** [2]

[1] **Prince Muqrin Chair - PMC for IT Security Technologies**
King Saud University
hsusanto@ksu.edu.sa

[2] **The Indonesian Institute of Sciences**
Information Security & IT Governance Research Group
heru.susanto@lipi.go.id

[2] **University of Brunei**
Information System Group
Susanto.net@gmail.com

*Abstract* – It is imperative for organizations to use Information Security Management System (ISMS) to effectively manage their information assets. ISMS starts with a set of policies that dictate the usage of computer resources. It starts with the "21 essential security controls" of ISO 27001, which give the basic standard requirements of information security management. Our research is concerned with the assessment of the application of these controls to organizations. STOPE (Strategy, Technology, Organization, People and Environment) methodologies were used to integrated domains as a framework for this assessment. The controls are mapped on these domains and subsequently refined into "246 simple and easily comprehended elements".

## I. INTRODUCTION

Information systems are becoming more complex as systems and the information processed, this has a huge effect on interfacing requirements, information storage and presentation formats and security. Along with the entry of the information age in every aspect of our life, such as organizations activities, business activities, transactional activities, the need for a standard set of information, especially concerning information security is eminent to ensure that the information delivered will be accepted or utilized as intended and to make sure the confidentiality and integrity of the information. Nowadays information is considered as an essential asset of any organization. As such, any security threat such as information theft, computer-assisted fraud, vandalism and computer hacking [5] must be seriously dealt with. Consequently, it is necessary to secure the interconnected business environments to protect information resource from potentials security threats. International Organization for Standardization (ISO) has issued an information security management system standard (ISMS) to bring information security under explicit management control [3]. Security management systems contain set of policies put place by an organization to maintain the security of their computer and network resources. These policies are based on the types of resources that need to be secured, depending on the organization. Some groups of policies can be applied to entire industries; others are specific to an individual organization [9]. To give organizations a starting point to develop their own security management systems, ISO and the IEC have developed a family of standards known as the Information Security Management System 27000 Family of Standards. This group of standards, starting with ISO 27001, provides organizations with the ability to certify their security management systems [9].

The paper is organized as follows. In the next section, we discuss the concept of STOPE methodology approach in information security. This is followed by a description, overview and refine of the research model and its hypotheses. The research method is then presented, followed by a discussion of the analysis and results. Finally, we discuss our finding and suggest future research.

## II. RELATED WORK

STOPE methodology schemes have been widely deployed to verify an information distributed activities on organization base on information security issues. In [2], Alfantookh proposed an approach for the assessment of the application of ISO 27001 essential information security controls. On STOPE point of view, Saleh et al [11] presented a novel approach methodology by introduced A STOPE model for the investigation of compliance with information security management standard ISO 17799-2005. It methodology makes it easy to implemented, measurable, and understandable. Bakry [5] also implemented STOPE by development of e-government.

## III. SECURITY MANAGEMENT STANDARD

ISO 27001 is designed to assure the confidentiality, integrity and availability of information assets, is exclusive to information security, and only addresses that issue [4]. The key areas identified by ISO 27001 for the implementation of an information security management system are:

- An information security policy
- Allocation of information security responsibilities within the organisation
- Asset classification and control
- Personnel security, responsibilities and training
- Physical and environmental security
- Communications and operational systems security
- Access controls

The decision as to what is appropriate depends upon understanding the risks and costs involved. Since risk appraisal includes all organizations and all departments, areas, staff and activities, the rationality and conformity of the appraisal is still a topic for research [7], [8]. Understanding the risk means knowing what the assets are, what the possible threats to those assets are, and the likelihood and possible impact of a security breach on the business. The goal of information security is to suitably protect this asset in order to ensure business continuity, minimize business damage, and maximize return on investments [6], [10], [12]. As defined by ISO 27001, information security is characterized as the preservation of CIA (Confidentiality, Integrity and Availability) [5]:

- **Confidentiality** – ensuring that information is accessible only to those authorized to have access.
- **Integrity** – safeguarding the accuracy and completeness of information and processing methods.
- **Availability** – ensuring that authorized users have access to information and associated assets when required.

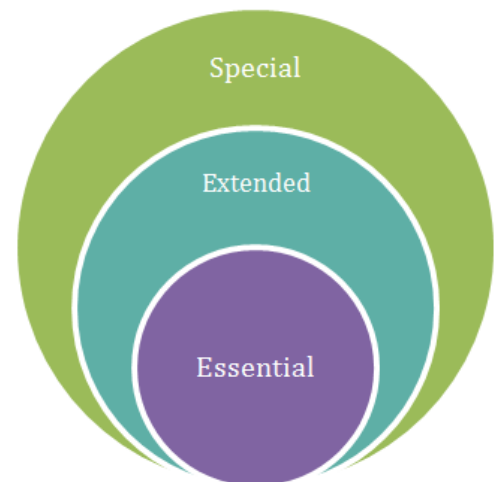

**Figure 1.** The three stages of security controls

The ISO 27001 standard contains 11 security control clauses, which are [3]:
- Security Policy *(contains of; 1 objective and 1 essential control)*;
- Organizing information security *(contains of; 2 objectives and 1 essential control)*;
- Asset Management security *(contains of; 2 objectives)*;
- Human Resources Security security *(contains of; 3 objectives and 3 essential controls)*;
- Physical and Environmental security *(contains of; 2 objectives)*;
- Communication and Operations Management *(contains of; 10 objectives)*;
- Access Control *(contains of; 7 objectives)*;
- Information Systems Acquisition, Development and Maintenance *(contains of; 6 objectives and 5 essential controls)*;
- Information Security Incident Management *(contains of; 2 objectives and 3 essential controls)*;

- Business continuity Management *(contains of; 1 objectives and 5 essential controls)*;
- Compliance *(contains of; 3 objectives and 3 essential controls)*;

STOPE is the abbreviation of the Strategy, Technology, Organization, People and Technology [5]. These five issues / STOPE called by domain of information security management systems. Every domain has several clauses, ejectives, controls including the essential ones [2].

## IV. STOPE METHODOLOGY

Bakry [4] introduced the basic elements of development in his STOPE profile as illustrated in *figure 2*. These elements are identified in the following:
- **Strategy:** the strategy of the country with regards to the future development of the industry or the service concerned.
- **Technology:** the technology upon which the industry or the service concered is based.
- **Organization:** the organizations associated with or related to, the industry or the service concerned.
- **People:** the people concerned with the target industry or service.
- **Environment:** the environment surrounding the target industry or service.

STOPE methodology separate analysis of the issue into five domains in the top. The separation is intended to better focus the analysis of existing problems with perspective Strategy, Technology, Organization, People and Environment. Table 1 illustrates structuring the ISO 27001 clauses and their "132" controls, including the "essential ones: 21 controls", over the five STOPE domains. The essential controls, which are concerned with the first security level of Figure 1, have been refined into 246 simple elements for the purpose of easing their assessment and application [1].

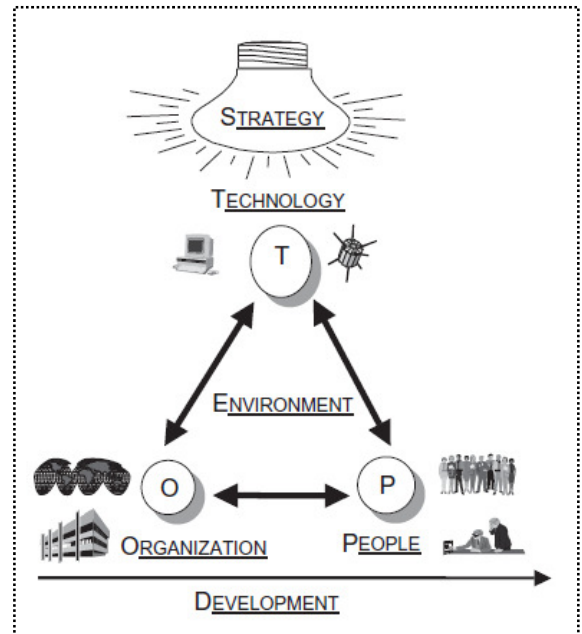

**Figure 2.** Bakry's STOPE Framework

## V. RE-REFINED ISO 27001

Refined is a deterministic process, which is ubiquitously present in the world. We verified and refined on standards existing; in order to determine the degree of clarity of each essential control over the parameters ISO 27001. Most organizations have a number of information security controls. Without a refined however, the controls tend to be somewhat disorganized and disjointed, having been implemented often as point solutions to specific situations or simply as a matter of convention. Sometimes a few essential controls are very difficult to understand, immeasurable and difficult to implemented, by organizations and stakeholders. Therefore we conducted a survey of respondents regarding the degree of clarity on the essential control. Refined perform by practitioners and experts of ISO 27001 information security field. We present refined in two domains of ISO 27001, namely Strategy and Technology domain. The results of our research, is seen in the explanation table below.

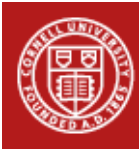

## VI. STRATEGY DOMAIN

Strategy domain comes from clause number 5 and section 5.1.1. of ISO 27001. It contains 1 objective and 2 controls, with 1 essential control on it, namely **Information security policy: document**, is a Single essential. In this strategy domain, respondents gave the feedback and agreed to all essential controls are carried by the ISO 27001, *table 1*. Clarity on the question and the type of answer each question did not lead to multiple interpretations. The table contains controls and questions that have been made refined from its original form in the manual books of ISO 27001. All respondents agreed with the refined alfantookh has done, that control, assessment issues and the question was quite reasonable and easily understood by stakeholders.

| Control | *An information security policy document should be approved by management, and published and communicated to all employees and relevant external parties* | | |
|---|---|---|---|
| Assessment Issue | **Refined Question** | **Status** | **Answer** |
| Existence | Is the information security policy document available? | approved | **Y or N** |
| Approval | Is the information security policy document approved by the management? | approved | **Y or N** |
| Publication | Is the information security policy document published? | approved | **Levels** |
| Internal communication | Is the information security policy document communicated to all ICT employees? | approved | **Levels** |
| | Is the information security policy document communicated to all ICT users? | approved | **Levels** |
| External communication | Is the information security policy document communicated to relevant external parties? | approved | **Levels** |
| Documentation | Is the reporting of the above exists? | approved | **Levels** |

**Table 1**: Assessing the control concerned with: "strategy: information security policy: *document*"

## VII. TECHNOLOGY DOMAIN

Technology domain comes from clause number 10, 11 and 12. It also ingredients of section number 12.2.1, 12.2.2, 12.2.3, 12.2.4, and 12.6.1 of ISO 27001. It contains 23 objectives and 73 controls, with 5 essential control on it, namely:
1. Input data validation *(table 2)*
2. Control of internal processing *(table 4)*
3. Message integrity *(table 5)*
4. Output data validation *(table 3)*
5. Control technical vulnerabilities *(table 6)*

Two refined and one additional question has been produced in this study, the addition of control (existence) and control of validation question. A question and grading system of answer that already exist can lead to multiple interpretations and ambiguous, added and refined question shown in the shaded tables.

| Control | *Data input to applications should be validated to ensure that this data is correct and appropriate.* | | |
|---|---|---|---|
| Assessment Issue | **Refined Question** | **Status** | **Answer** |
| Existence | Plausibility checks exist to test the output data reasonability? | added | **Y or N** |
| Validation | Is the examination for the input business transaction, standing data and parameter tables exist? | modified | **Levels** |
| | Is the automatic examination exists? | approved | **Y or N** |
| | Is the periodic review and inspection available? | modified | **Levels** |
| | Is the response of procedures to validation exist? | approved | **Y or N** |
| Management | Is the logging of events exists? | approved | **Y or N** |
| Accountability | Are the responsibilities defined? | approved | **Levels** |
| Documentation | Is the reporting of the above exists? | approved | **Levels** |

**Table 2:** Assessing the control concerned with "technology: information systems acquisition, development and maintenance: correct processing in applications: *input data validation"*

| Control | Data output from an application should be validated to ensure that the processing of stored information is correct and appropriate to the circumstances. | | |
|---|---|---|---|
| Assessment Issue | Refined Question | Status | Answer |
| Existence | Plausibility checks exist to test the output data reasonability? | added | **Y or N** |
| Validation | Is the provided information for a reader or subsequent processing system sufficient to determine the accuracy, completeness, precision and classification of the information? | modified | **Levels** |
| | Is the periodic inspection exists? | approved | **Levels** |
| | Are the responding procedures validation test exist? | added | **Y or N** |
| Practice | Is the checking that programs are run in order exists? | approved | **Y or N** |
| | Is the checking that programs are run at the correct time exists? | approved | **Y or N** |
| Accountability | Are the responsibilities defined? | approved | **Y or N** |
| Documentation | Is the reporting of the above exists? | approved | **Levels** |

**Table 3:** Assessing the control concerned with "technology: information systems acquisition, development and maintenance: *output data validation"*

Internal control processing, message integrity and Control technical vulnerabilities, the respondents agreed that has been done alfantookh refined. no improvement and rebuttal to control, the *table 4, table 5, table 6.* Thus alfantookh refined enough to be accepted by stakeholders ISO 27001.

| Control | Validation checks should be incorporated into applications to detect any corruption of information through processing errors or deliberate acts. | | |
|---|---|---|---|
| Assessment Issue | Refined Question | Status | Answer |
| Validation | Is the validation of generated data, or software, exists? | approved | **Y or N** |
| | Is the validation of downloaded data, or software, exists? | approved | **Y or N** |
| | Is the validation of the uploaded data, or software, exists? | approved | **Y or N** |
| Protection | Is the use of programs that provide recovery from failure exists? | | |
| | Is the termination of programs at failure exists? | approved | **Y or N** |
| | Is the protection against attack exists? | approved | **Y or N** |
| Practice | Is the checking that programs are run in order exists? | approved | **Y or N** |
| | Is the checking that programs are run at the correct time exists? | approved | **Y or N** |
| | Is the checking that programs terminate correctly exists? | approved | **Y or N** |
| | Is the logging of events exists? | approved | **Y or N** |
| Accountability | Are the responsibilities defined? | approved | **Y or N** |
| Documentation | Is the reporting of the above exists? | approved | **Levels** |

**Table 4:** Assessing the control concerned with "technology: information systems acquisition, development and maintenance: *control of internal processing"*

| Control | Requirements for ensuring authenticity and protecting message integrity in applications should be identified, and appropriate controls identified and implemented. | | |
|---|---|---|---|
| Assessment Issue | Refined Question | Status | Answer |
| Requirements | Are message integrity requirements specified? | approved | **Levels** |
| Protection | Are message integrity protection measures implemented? | approved | **Levels** |
| | Implemented protection measures are suitable to message integrity requirements | approved | **Levels** |
| Practice | Is the logging of events exists? | approved | **Y or N** |
| Accountability | Are the responsibilities defined? | approved | **Y or N** |
| Documentation | Is the reporting of the above exists? | approved | **Levels** |

**Table 5:** Assessing the control concerned with "technology: information systems acquisition, development and maintenance: *message integrity*

| Control | *Timely information about technical vulnerabilities of information systems being used should be obtained, the organization exposure to such vulnerabilities evaluated and appropriate measures taken to address the associated risk.* | | |
|---|---|---|---|
| Assessment Issue | **Refined Question** | **Status** | **Answer** |
| Inventory of technical assets | Are the technical specifications of systems and their components exist? | approved | **Y or N** |
| Vulnerability | Are the vulnerabilities of technical assets identified? | approved | **Levels** |
| | Are the risks associated with vulnerabilities identified? | approved | **Levels** |
| Protection | Protection measures that respond to risks are identified | approved | **Levels** |
| | Are the protection tools evaluated before use? | approved | **Levels** |
| | Is the awareness on potential vulnerabilities among the right people exists? | approved | **Levels** |
| Practice | Does the monitoring to manage problems exist? | approved | **Y or N** |
| | Are logging of events exist? | approved | **Levels** |
| Accountability | Do the defined responsibilities exist? | approved | **Y or N** |
| Documentation | Is the reporting of the above exists? | approved | **Levels** |

**Table 3:** Assessing the control concerned with "technology: information systems acquisition, development and maintenance: *control of technical vulnerabilities"*

A computer tool which eases the assessment process, as a assessment tool, was developed simultaneously. However by computer tool mentioned, assessment process of the organization is automated. The final result of calculation might be easily analyses, shorter and precision, done with the developed computer tool.

## VIII. CONCLUSION REMARKS

The work presented in this report included refining the ISO/IEC 27001 21 essential controls using STOPE approach. The refinements were based on previous work, but it has been validated by experts from Prince Muqrin Chair (PMC) for information security. In addition, our designed computer tools, which ease the process of assessment of the implementation of security standards, are also assessed. Next step after our work is to refine and validate the other controls for the ISO 27001 (special and extended) and integrate them in the computer tool, to complete the whole information security standards for assessment according to ISO 27001.

## IX. ACKNOWLEDGEMENT

This project is under supported of Prince Muqrin Chair (PMC) for Information Technologies Security, Information Security Management Systems Research Group. King Saud University. Thanks to Prof. Dr. Saad Hajj Bakry for supervised us during project.